\def\mbf#1{\bm{\mathrm{#1}}}
\def\mrm#1{\mathrm{#1}}

\documentclass [prb,superscriptaddress]{revtex4}
\usepackage {graphicx}
\usepackage {amsmath}
\usepackage {bm}
\usepackage {hyperref}
\usepackage {citesort}

\begin{document}


\title{
Spatial Coherence of Synchrotron Radiation
}

\begin{abstract}
Theory and measurement of spatial coherence of synchrotron radiation beams are 
briefly reviewed. Emphasis is given to simple relationships
between electron beam characteristics and far field properties of the light 
beam. 
\end{abstract}

\author{R. Co\"\i sson}
\affiliation{Dipartimento di Fisica and INFM, Universit\`a di Parma, 
43100 Parma, Italy}

\author{S. Marchesini}
\affiliation{
Physics \& Advanced Technologies, Lawrence Livermore 
National Laboratory,
Livermore, California 94550}


\date{\today}
\maketitle

\section*{Introduction}

Synchrotron Radiation (SR)\cite{handbook, esrfcd, wordscientific} has been widely used since the 80's as a tool for 
many applications of UV, soft X rays and hard X rays in
condensed matter physics, chemistry and biology.
The evolution of SR sources towards higher brightness has led to the design of 
low-emittance electron storage rings (emittance is the product
of beam size and divergence), 
and the development of special source magnetic structures,
as undulators. This means that more and more photons are
available on a narrow bandwidth and on a small collimated beam; in other words
there is the possibility of getting a high power in a coherent
beam.
In most applications, a monochromator is used, and the temporal coherence of the
light is given by the monochromator bandwidth. With
smaller and smaller sources, even without the use of collimators,
the spatial coherence of the light has become appreciable, first in the UV
and soft X ray range, and then also with hard X rays. This has made possible new
or improved experiments in interferometry, microscopy,
holography, correlation spectroscopy, etc.
 \cite{attwood,cornacchia,attwood_pt,attwoodbook,kondratenko,alferov,tang}.
In view of these recent possibilities and applications, it is useful to review
some basic concepts about spatial coherence of SR, and its
measurement and applications.
In particular we show how the spatial coherence properties of the radiation in
the far field can be calculated with simple operations from the
single-electron amplitude and the electron beam angular and position spreads. 
The gaussian approximation will be studied in detail for a
discussion of the properties of the far field mutual coherence and the estimate
of the coherence widths, and the comparison with the
VanCittert-Zernike limit. 

\section{Spatial Coherence (SC)}

First let us remind some concepts and define some symbols about SC of a
quasi-monochromatic field in general.
If we have paraxial propagation of a random electromagnetic field
$f(\mbf{p})$ (where
$\mbf{p}=(x,y)$ is a point in the transverse plane) along a direction
$z$, the filed
$f( \mbf p)n$ at $z=0$ propagates in the Fresnel approximation

\begin{equation}
\label{eq:1}
 f_z(\mbf p)=\frac 1  {(\lambda z)^2}  \int{f_0(\mbf p_0)\,
e^{-\frac {ik}  z (\mbf p-\mbf p_0)^2} \,d^2 \mbf p  }\,,
\end{equation}
\noindent
and in the Far Field (FF) (Fraunhofer region), where most observations
are done, 
we have
\begin{equation}\label{eq:3}
 \tilde f (\mbf k)=\mrm F_{\mbf p\rightarrow \mbf k}f(\mbf p)=\frac 1
{\sqrt{2\pi}} \int {f(\mbf p)\, e^{i\mbf k \cdot \mbf p}\,d^2\mbf p}
\end{equation}

where we have dropped the index $0$ and indicated with $F_{\mbf p \rightarrow \mbf k}$
the Fourier transform operator from $\mbf p$ to $\mbf k$
domain.
Here we use as a variable the transverse component of the wavevector
$\mbf k=(k_x,k_y)$, the observation angle is then
\begin{equation} \label{eq:2}
 \theta= \mbf k \frac{\lambda} {2\pi} \,,
\end{equation}
\noindent
According to eq. \eqref{eq:3}, angles are expressed in terms of reciprocal space
coordinates, as is natural in diffraction optics. \par

``Second-order'' statistical properties of the field are described by the
``mutual intensity" (m.i.) (see \cite{goodman}), i.e. the ensemble average
of the products of fields at two points.

It is convenient to express the m.i. as a function of the average and difference
coordinates: $\mbf p_1=\mbf p-\Delta \mbf p/2$, $\mbf p_2=\mbf
p+\Delta \mbf p/2$ and, in reciprocal space, $\mbf k_1=\mbf k-\Delta \mbf k/2$,
$\mbf k_2=\mbf k+\Delta \mbf k/2$ \par
\begin{equation}
\mrm M f(\mbf p,\Delta \mbf p) = \langle f^*(\mbf p-\Delta \mbf
p/2)f(\mbf p+\Delta \mbf p/2\rangle
\end{equation}
For simplicity we will also use these symbols:
$$\mrm I f(\mbf p)=\mrm M f(\mbf p,0)$$ the intensity
and
$$\mrm {C} f(\Delta \mbf p) = \int Mf(\mbf p, \Delta \mbf p) d^2 \mbf p$$
 the (integrated) autocorrelation.
The degree of (spatial) coherence is defined as
 \begin{equation}
\mu f(\mbf p, \Delta \mbf p)\equiv \mrm{M} f(\mbf p ,\Delta \mbf p)/ \sqrt{I 
(\mbf p -\Delta \mbf p/2)
I (\mbf p+\Delta \mbf p/2)}
\end{equation}
The Fresnel propagation of the m.i. can be expressed as: \begin{equation}
\label{fresnelmi} \mrm M f_z(\bar{\mbf p},\Delta \bar{\mbf
p})=\frac{1}{(\lambda z)^2} \int \mrm M f_0(\mbf p,\Delta \mbf p )e^{i\frac k
{z}(\bar {\mbf p} - \mbf p) (\Delta \bar{\mbf p}-\Delta \mbf p )}d^2
\mbf p
\,d^2 \Delta \mbf p
\end{equation}
and in the FF
\begin{equation}
\left \langle \tilde f(\mbf k-\tfrac {\Delta \mbf k} 2)\tilde f (\mbf
k+\tfrac {\Delta \mbf k} 2)\right \rangle =
\int \mrm M f(\mbf p,\tfrac {\Delta \mbf p} 2)
 e^{i \mbf p  \cdot \Delta \mbf k+\mbf k \cdot \Delta \mbf p}
d^2 \mbf p \, d^2\Delta \mbf p
\end{equation}
or, with our simplified notation,
$$
 \mrm M \tilde f (k, \Delta k) = \mrm F_{\mbf p\rightarrow \Delta
\mbf k}
F_{\Delta \mbf p \rightarrow \mbf k} \mrm M f(\mbf p,\Delta \mbf  p)\,.
$$
From this, two useful reciprocity relations connecting source and FF
intensity/coherence properties can be derived 
\cite{ friberg, coissoncern}:
\begin{equation}
\label{eq:8}
 \mrm {FC} f (\mbf k) = \mrm {IF} f(\mbf k)
\end{equation}
\begin{equation}
\label{eq:9}
\mrm{ FI} f(\Delta\mbf  k) = \mrm CFf(\Delta\mbf  k)
\end{equation}
and reciprocal ones interchanging source and FF.
Properties of non-stationary random functions can also be described by
the Wigner function (WF) (which is a photon number distribution in phase 
space, if divided by $\hbar\omega$)
\cite{kim,coissonspie86,walther} :
\begin{eqnarray}
\nonumber
 \mrm{W} f(\mbf p ,\mbf k) &=&
\int \left \langle f(\mbf p-\Delta \mbf{p} /2) f^*(\mbf {p}+\Delta
\mbf {p} /2)
\right \rangle  e^{i\Delta \mbf{p} \cdot \mbf k } d^2\Delta \mbf p  \\
&=&
\int  \left \langle \tilde
f (\mbf k-\Delta \mbf k /2) \tilde f^*(\mbf k+\mbf k /2) \right
\rangle  e^{i \Delta \mbf k \cdot \mbf p} d^2\Delta \mbf k
\end{eqnarray}
In fact from the definition we see that Fourier-transforming the WF with respect
to k one gets the m.i. of f(x), while transforming with respect to x
gives the m.i. of $\tilde f(k)$. We also remind that the intensity at the object plane $If=\int Wf dk$ and in the far field
$I\tilde f = \int Wf dx $.

 An equivalent description, with essentially the same
characteristics,
could be obtained with the Ambiguity function \cite{papoulis:josa}
\begin{equation} Af(\Delta \mbf p ,\Delta \mbf k) = \int <f(\mbf p-\Delta \mbf p
/2) f^*(\mbf p+\Delta \mbf p /2)> e^{i \mbf p \cdot \Delta \mbf k}\, d^2\mbf p
\end{equation} Both Wigner and Ambiguity functions are real (almost
always positive) functions and can be considered as a phase space energy
density: notice that this phase space area is dimensionless. $Wf$
propagates in the same way of the ``radiance" (or ``brightness") of geometrical
optics:

\begin{equation}
\label{fresnelwf}
Wf_z(\mbf p,\mbf k )=Wf_0(\mbf p-\frac {\mbf k} k z,\mbf
k)\,,
\end{equation}
and the same for $Af_z(\Delta \mbf p, \Delta \mbf k)$.

\subsection{Gaussian approximation}

A gaussian model (also called a gaussian Schell model\cite{mandelandwolf}) of a partially coherent
field has a radiance which has a 4-D gaussian distribution in phase space. 
From now on let us for simplicity consider one transverse dimension, say x 
(and $k_x$ will be called k for short):
we have then [ $\mrm Wf$ or $\mrm Af$ ] of the form (using our previous 
symbols):
\begin{equation}
\mrm W f(x, k) = N_1 \exp \left (- \tfrac 1 2 \tfrac {x^2} {\sigma_I^2}\right)
\exp \left (-\tfrac 1 2 \tfrac{k^2} {s_I^2}\right)
\end{equation}

the $\mrm MI$ is then
\begin{equation}\label{eq:MI} 
Mf(x,\Delta x) = 
N_2 \exp \left (-\tfrac 1 2 \tfrac {x^2} {\sigma_I^2}\right)
\exp \left (-\tfrac 1 2 \tfrac {\Delta x^2} {\sigma_M^2}\right)
\end{equation}
where
\begin{equation}
\sigma_M=\frac 1 {s_I}
\end{equation}
(in agreement with eq. \ref{eq:8}), if we define $\sigma_M$ as the 
width of $Cf(\Delta x)$. Here we have indicated with N the normalisation constants)

This MI clearly satisfies separability between x and $\Delta x$
(Walther's condition \cite{walther}).
When $\sigma_M\ll \sigma_I$ we have the ``quasi-homogeneous" approximation ($Mf(x,\Delta x)=I f(x)\mu f(\Delta x)$), and
the factor function of $x$ has the meaning of the
intensity \cite{carter}.

We easily see that this model satisfies the Schell condition (that's why it is also called ``gaussian Schell'' model) that the degree of coherence depends only on the separation between two points $\Delta x$) 
: eq. \ref{eq:MI} can be
written:
\begin{equation}
 \mrm M f(x,\Delta x) =
N \exp \left \{-\tfrac 1 2 \frac{(x-\Delta x/2)^2}{2\sigma_I^2} \right \}
 \exp\left \{-\tfrac 1 2 \frac{(x+\Delta x/2)^2}{2\sigma_I^2} \right \} 
\exp \left \{-\tfrac 1 2  \frac{\Delta x^2}{\sigma_\mu^2} \right \}
\end{equation}
where
\begin{equation}
\label{eq:17}
 \frac{1}{\sigma_\mu^2} = \frac{1}{\sigma_M^2}-\frac{1}{4\sigma_I^2}
\end{equation} 
In particular, we see that for a perfectly coherent gaussian beam, $\sigma_M = 
2\sigma_I$. The Schell and Walther conditions are satisfied simultaneously only for a plane wave and gaussian wave: writing the two conditions,
\begin{equation}
\label{eq:waltherandshelltogauss}
\left [ \mrm I  f \left (x+\tfrac{\Delta x} {2} \right ) 
\mrm I f \left (x-\tfrac{\Delta x} {2} \right )\right ]^{\frac 1 2}
    \mu f  (\Delta x )=
\mrm M  f \left (x,\Delta x \right )=
\mrm I  f \left (x \right ) 
\mrm m f ( \Delta x )
\end{equation}
If we apply the logarithm and call 
$h(\Delta x )=
\log [ \mu f ( \Delta x )/\mrm m f ( \Delta x )]
$, $L(x)=\log [\mrm I f(x)]$ Eq. \ref{eq:waltherandshelltogauss} becomes:
$$
2L(x)-L\left (x+\tfrac{\Delta x} {2}\right )
-L\left (x-\tfrac{\Delta x} {2}\right)=2h(\Delta x)
$$
By Taylor expanding $L$ we see that in order for the left term to be
 dependent only on $\Delta x$, terms higher than 2 must be 0, i.e. a 
Gaussian, exponential or flat intensity only.

\section{Mutual intensity of Synchrotron Radiation}

A characteristic of SR is that it is the random superposition of a large number
of rather collimated elementary waves emitted by each electron
of the beam \cite{kim, coissonspie86, coisson:jsr}. Let us call $\tilde a (\mbf k ) $ the well-known far-field amplitude
(or square root of the intensity) emitted by a single electron. It can
be seen as the FT of the amplitude at the source $a(\mbf p )$, which of course is
not a Dirac delta because of the diffraction corresponding to
the limited angular aperture (and this is a limit to the possibility of
localizing an electron by observing or imaging the emitted SR). The electron
beam is characterized by a transverse spatial distribution g(\mbf p) and an angular
distribution $\gamma (\mbf k)$, which are to a good
approximation both gaussian. The ratio of the beam size and angular aperture is
called the beta function and it is known from the machine
physics. Usually the source is in a place where position and angular
distribution are uncorrelated; otherwise it is possible to define an effective
source position at the ``waist" point where the two distibutions are 
uncorrelated. The ``waist" points may be 
different for the vertical and the horizontal distributions.
 
We will consider for simplicity one transverse coordinate, say $x$ and $k$.  
The superposition of all elementary contributions can be best described 
in phase space, where the
Wigner function (or Ambiguity function) can be obtained by a convolution 
of the two distributions (electron, and single-electron light)
\cite{kim,coissonspie86}.

\begin{equation}
\label{eq:19}
\mrm W f(x,k)=g(x)\gamma (k) ** \mrm Wa(x,k)
\end{equation}

where $**$ indicated the convolution with respect to both variables.
The source and FF mutual intensities are then \cite{coisson:ao95}: 

\begin{equation}
\label{eq:20}
 \mrm Mf(x,\Delta x) = \tilde\gamma (\Delta x)[g(x)*\mrm
Ma(x,\Delta x)]
\end{equation}
 and
\begin{equation}
\label{eq:21}
 \mrm M\tilde f (k, \Delta k) = \tilde g (\Delta k) [\gamma
(k)*\mrm M\tilde a (k,\Delta k)] 
\end{equation}
In particular, for the FF intensity:
\begin{equation}
\mrm I\tilde f (k) = \gamma (k)* \mrm I\tilde a (k) 
\end{equation}
In order to give estimates of sizes and correlation distances of SR, it is
useful to use a gaussian approximation for the SR distributions $a(x)$ and 
$\tilde a (k)$. 
Actually they are not gaussians, but this
approximation is rather good for two reasons: $g(x)$ and $\gamma
(k)$ being gaussians, the convolution is close to a gaussian, except on the 
tails (as $a(x)$ has
long tails), and the part that is used is just the central one.

With this gaussian approximation, the source and FF are characterised by 6 
gaussian widths. 

Let us call $\sigma_I$ the characteristic width of the intensity (so that
$I(x)=\exp(-x^2/2\sigma^2)$) at the source, and $s_I$ the FF intensity width.
The M.I. of the source is given by eq. \ref{eq:MI} 

As we have seen (eq. \ref{eq:MI}), the degree of coherence $\mu f(x)$ has a width which 
is related to the other two widths by:
\begin{equation}
 \frac{1}  {\sigma_\mu^2} = \frac 1 {\sigma_M^2} - \frac 1
{4\sigma_I^2}
\end{equation}

And analogously, if we use $s$ for the FF widths:
\begin{equation}
 \frac{1}{s_\mu^2} = \frac 1 {s_M^2} - \frac 1 {4s_I^2}
\end{equation}

On the other hand, if we apply the reciprocity relations 
(Eq. \ref{eq:8}, \ref{eq:9} ) to the 
gaussian case, we have:
\begin{equation}
\label{eq:reciprocity}
 s_M = \frac 1 {\sigma_I}\,,
\text{ and }
s_I = \frac 1{\sigma_M}
\end{equation}

\begin{figure*}[htbp]
\centerline{
\includegraphics[width=2in]{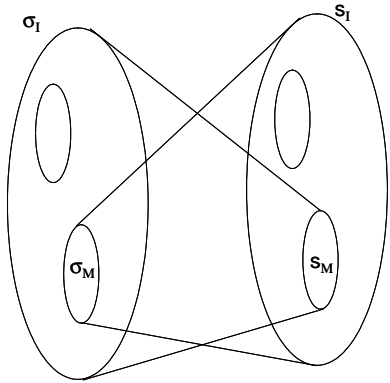}
}
\caption{
Illustration of the source - far field reciprocity relations 
(eq. \ref{eq:reciprocity}), note that in
the gaussian case the ratio between beam width and coherence width is the same in the near field and far field (as well as in all sections
in between \cite{coisson:jsr})
}
\label{fig:1}
\end{figure*}

This is illustrated in fig \ref{fig:1}

We want now to correlate these widths with the electron and SR
characteristics.  We approximate the single-electron FF amplitude with
\begin{equation}
\label{eq:26}
\tilde a (k) =  \exp \left ( -\frac{k^2}{4\rho^2} \right)
\end{equation}
In this way we have defined $\rho$ as the gaussian width of the FF intensity. 
The angular width $\rho$ is of the order of the relativistic factor of
the electrons 
(multiplied by $2\pi/\lambda$, in our reciprocal space
units) \cite{alferov, coisson:oe, coisson:ao88}

If we also apply to the gaussian case eqs. \ref{eq:19},
we get
\begin{equation}
 s_I^2 = s_e^2 + \rho^2
\text{, and }
 \sigma_I^2 = \sigma_e^2+1/4\rho^2
\label{eq:27}
\end{equation}
Putting together these relations, we can eventually determine the intensity and 
coherence properties of the FF as a function of electron beam
(and single-electron radiation) data:

 \begin{eqnarray}
\nonumber
s_I  =(s_e^2 +\rho^2)^{1/2}\,,\,
s_M  = (\sigma_e^2 +1/4\rho^2)^{-1/2} \\
s_\mu  = \left (\sigma_e^2+\frac 1{4\rho^2} -\frac 1{4(s_e^2+\rho^2)} \right)^{-1/2}
\label{eq:28}
 \end{eqnarray}

In the perfectly coherent limit ($s_e<<1/\rho$ and $s_e<<\rho$) we have 
$s_\mu=\infty$, $s_I=\rho$ and $s_M=\rho$.
The quasi-homogeneous case is when $s_e>>\rho$ and $\sigma_e>>1/\rho$: 
in this case 
\begin{equation}
s_\mu = (\sigma_e^2+1/4\rho^2)^{-1/2} \simeq 1/\sigma_e
 \end{equation}
This result coincides with the VanCittert-Zernike theorem, 
valid in the limit of a completely incoherent source.  
In general, however (for a rather coherent
beam, that is a beam produced by an electron beam with small $\sigma_e$ and 
$s_e$), the VanCittert-Zernike theorem  needs a correction \cite{coisson:ao95}.

It may also be of interest to know the resolution for imaging the source 
on the basis of FF intensity and coherence measurements. 
In principle, we can get both $\sigma_e$ and $s_e$ by measuring $s_I$ 
and $s_M$ or $s_\mu$: 
from the previous equations we see that 
from eq \ref{eq:27} we get 
\begin{eqnarray}
\nonumber
 s_e^2&=&s_I^2-\rho^2 \,,\\
 \sigma_e^2&=&1/s_M^2-1/4\rho^2 \,.
\label{eq:30}
\end{eqnarray}
However in practice the low precision of correlation measurement with the 
unfavorable propagation of errors, 
makes the method usable only if $4\rho^2/s_M^2-1$ and $s_I^2/\rho^2-1$ are not 
much smaller than one, i.e. the beam is not much smaller than the diffraction 
limit). 

In these remarks we have considered always a quasi-monochromatic component of 
the field; in other words we imagine 
the light to be filtered before by a monochromator. It may be worthwhile to 
mention that SR, and in particular the radiation from undulators, 
is not ``cross-spectrally pure" as defined by Mandel \cite{mandel}, as the 
spectrum depends on angle, and then the spatial coherence 
and spectral characteristics cannot be separated, a subject that has not yet 
been analysed in the literature.  

\section{Effects of quality of optical elements.}

In recent machines where spatial coherence becomes appreciable over a fraction 
of the photon beam width, 
or in other words is very well collimated (near the diffraction limit), 
the effect of imperfection of optical elements, as mirrors \cite{coissonreport,wang}
 or Berillium  windows \cite{snigirev:nima}
 strongly influences the beam quality. 
For mirrors, if the rms slope error is $\varsigma$, this must be compared with 
$\theta_{coh}=\lambda s_\mu /2\pi$:
 in order to have small distortions we should 
have $\varsigma<<\theta_{coh}$
For windows, a uniform illumination will become non-uniform, with a contrast
$$ C=2\pi h/\lambda (n-1)$$.

Some authors have called this degradation of beam quality  a ``reduction of 
coherence" \cite{snigirev:nima, snigirev:rsi}. Actually this is not 
precise  \cite{nugent:optexp}, as the speckle-like field produced by a 
random deflection from a rough surface
(or refraction from a rough window) is still capable of producing
interference fringes in a Young experiment if the original wave was spatially
coherent. In fact, the optical path (as a function of x,y) is fixed in
time, it is a single realization of a random function, in other words a
deterministic function (although not known in detail). We have to
distinguish averages in time from averages over an ensemble
of optical elements with similar statistical properties. The measure of 
correlation distance is given by 
$\sigma_\mu$ or $s_\mu$, not by $\sigma_M$ or $s_M$, as the latter ones maybe 
short, for example, in a perfectly coherent light with strong and rapid 
spatial variations of intensity. 

In other words, coherent light stays coherent, even after passing through
 a random media. The photons in a coherent volume in phase space never mix
 with others as a consequence of the Liouville theorem. However when 
we perform a measurement,  we normally measure projections (intensity) 
or slices (interferometry) in phase space. 
In the case of a Young's slits experiment for example, the two slits
act as slices in the phase space, the beams diffracted from the slits have 
lost directionality, and different volumes in phase space are therefore mixed.
In a intensity interferometry experiment, we integrate the phase space distribution over the angles \cite{vartanyants:optcom}. 

\begin{figure*}[htbp]
\centerline{
\includegraphics[width=7in]{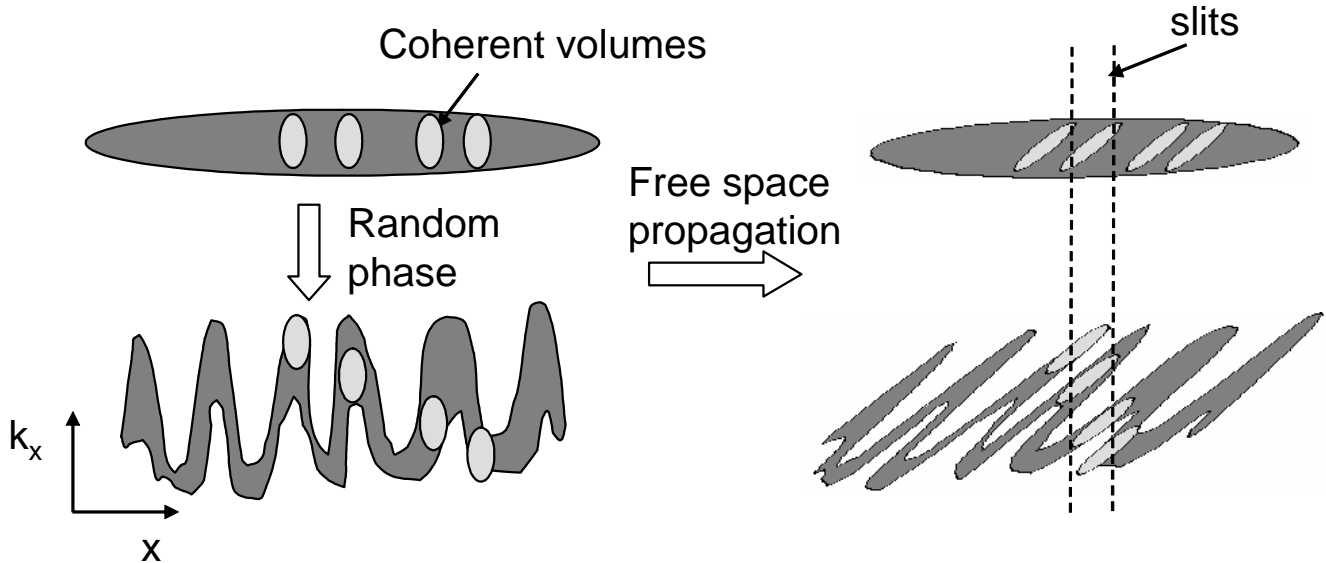}}
\caption{propagation of the wigner function: (top) a gauss-shell beam  
propagates in free space, and a coherent volume is selected by two slits. 
(bottom) the same beam after passing through a random phase object 
\label{fig:2}}
\end{figure*}

\section{Measurements}
The first soft x-ray interferometric measurements 
with synchrotron radiation were performed by Polack
 et al \cite{polak} using two mirrors with an angle 
between them of 2.25 arcmin at 6$^0$ grazing angle. 
Coherence measurements using Young slits have been performed 
by many groups in the soft X-ray range \cite{takayama, xu, chang, paterson}. 
Takayama used a young-slit experiment  to characterize 
the emittance of the electron beam \cite{takayama1}. 

\begin{figure*}[htbp]
\centerline{
\includegraphics[width=3in]{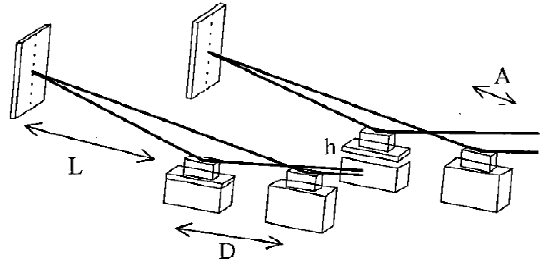}}
\caption{experimental setup used to perform hard x-ray interferometric 
characterization of the coherence.  By moving $D$ or changing the angle
 of incidence, or the height $h$ of one mirror one can study the vertical 
coherence, while by tilting one mirror it is possible to study the
 horizontal coherence \cite{fezzaa}. 
\label{fig:3}}
\end{figure*}

In the hard x-ray the first interferometric measurement of the beam coherence 
was performed using two mirrors at grazing incidence acting as
 slits \cite{fezzaa,marchesini:oe96,marchesini} (Fig. \ref{fig:3}). 
Normal slits have also been applied \cite{leitenberger}.

Other measurements of  coherence have been performed by diffracting 
x-rays from  a wire \cite{snigireva,kohnprl}, using Talbot effect 
\cite{cloetens},  a mask of coded apertures called a uniformly redundant array
 (URA) \cite{lin:prl}. Other techniques include using nuclear resonance 
from a rotating disk and measuring the spatial coherence in the time domain 
(the rotating disk acts as a 'prism' of increasing angle) \cite{baron}, and
intensity interferometry  \cite{yabashi:2001}. The latter has been used 
to measure the spatial as well as longitudinal 
coherence \cite{yabashi:2002} and characterize the 3 dimensional 
x-ray pulse widths. 
Variation of the visibility of a speckle pattern can also be used as an 
indication of the coherence width \cite{abernathy}.

\begin{acknowledgments}
 This work was performed under the auspices of the U.S. Department 
of Energy by the Lawrence Livermore National Laboratory under Contract No. 
W-7405-ENG-48 and the Director, Office of Energy Research, Office of Basics 
Energy Sciences, Materials Sciences Division of the U. S. Department of 
Energy, under Contract No. DE-AC03-76SF00098.
\end{acknowledgments}


\begin{thebibliography}{17}
\bibitem{handbook} see for ex. {\it Handbook on Synchrotron
Radiation}, E. E. Koch, Ed.,
Vol.1-4, North Holland, Amsterdam (vol.1 in 1983).    

\bibitem{esrfcd} Synchrotron Light CD-ROM, ISBN 3-540-14888-4, IMEDIASoft/Springer-Verlag/ESRF, 2000

\bibitem{wordscientific} Series on Synchrotron Radiation Techniques and
Applications, Word Scientific, Singapore 2000
Volume 1:
Synchrotron Radiation Sources - A Primer, edited by Herman Winick
Volume 5
Synchrotron Radiation Theory and Its Development, edited by Vladimir A 
Bordovitsyn
Volume 6 Insertion Devices for SR and FEL by F Ciocci, G Dattoli, A Torre \& A 
Renieri

\bibitem{attwood} D. Attwood, K. Halbach, K.-J. Kim, ``Tunable coherent X-rays", Cambridge University Press; (September 1999)

\bibitem{cornacchia} M. Cornacchia, H. Winick, XV Int. Conf. on High
Energy Accel., Hamburg 1992.
Science \underbar{228},1265-72 (1985). \par
\bibitem{attwood_pt} D. Attwood, ``New opportunities at soft X-ray wavelengths",Phys. Today, August 1992, p.24-31. 
\bibitem{attwoodbook} ``Soft X-Rays and Extreme Ultraviolet Radiation : Principles and Applications'', D. Atwood, Cambridge University Press, (Cambridge 1999).
\bibitem{kondratenko} A. M. Kondratenko, A. N. Skrinsky ``Use of radiation of
electron storage
rings in X-ray holography of objects", Opt. Spektrosk. \underbar{42}, 338-344
(1975); Engl. transl. in: Opt. Spectrosc. \underbar{42}, 189-192 (1977). \par

\bibitem{alferov} D. F. Alferov, Yu. A. Bashmakov, E. G. Bessonov, ``Theory of
undulator
radiation", Zh. Tech. Fiz. \underbar{48}, 1592-1597 and 1598-1606 (1978); Engl.
transl. in: Sov. Phys. Tech. Phys. \underbar{23}, 902-904 and 905-909 (1978).

\bibitem{tang} E.Tang, P.Zhu, M.Cui, "Coherence mode of SR", 
Acta Optica Sinica 18, 1645 (1998).

\bibitem{goodman} J. W. Goodman, {\it Statistical Optics}, ch. 5, J.Wiley, NY
1985.   


\bibitem{friberg}  A. Friberg, E. Wolf, ``Reciprocity relations with
partially coherent sources",  Opt. Acta \underbar{30}, 1417-1435(1983). 

\bibitem{coissoncern} R. Co\"\i sson, ``Source and far field coherence
functions", Note SPS/ABM/RC 81-11, CERN, Geneva 1981.  

\bibitem{kim}  K.-J. Kim, ``A new formulation of synchrotron radiation
optics using the
Wigner distribution", Proc SPIE \underbar{582}, 2-9 (1986).      

\bibitem{coissonspie86}  R. Co\"\i sson, R. P. Walker, ``Phase space
distribution of brilliance of undulator sources"  Proc. SPIE \underbar{582},
24-29 (1986). \par

\bibitem{walther} A. Walther, ``Radiometry and coherence'', J. Opt. Soc. Am.
\underbar{58},1256-59 (1968).  

\bibitem{carter}  W. H. Carter, E. Wolf, ``Coherence and radiometry
with quasihomogeneous planar sources", J. Opt. Soc. Am. \underbar{67},785
(1977); 

\bibitem{papoulis:josa} A. Papoulis, ``Ambiguity function in Fourier
optics", J. Opt. Soc. Am.\underbar{64}, 779-788 (1974).    \par

\bibitem{mandelandwolf}
L.Mandel and E.Wolf, Optical Coherence and Quantum Optics, (Cambridge University Press, Cambridge, 1995)



\bibitem{coisson:jsr} R. Co\"\i sson, S. Marchesini,
Gauss-Shell sources as model for Synchrotron Radiation, Journal of Synchrotron Radiation 4(5), 1997, 263-266.



\bibitem{coisson:ao95} R. Co\"\i sson, ``Spatial coherence of synchrotron radiation", Appl. Opt. 34, 904-8 
(1995)

\bibitem{coisson:oe} R. Co\"\i sson, ``Effective phase space widths of
undulator radiation", Opt. Eng. \underbar{27}, 250-52 (1988).     

\bibitem{coisson:ao88} R. Co\"\i sson, B.Diviacco, ``Practical estimates
of peak flux and brilliance of undulator radiation on even harmonics", Appl.
Opt. \underbar{27},1376-7 (1988).     \par

\bibitem{mandel} L. Mandel, ``Concept of cross-spectral purity in
coherence theory", J. Opt. Soc. Am. \underbar{51}, 1342 (1961).

\bibitem{coissonreport} R.Co\"\i sson, ``Estimation of the effect of
slope errors on soft X-ray optics", report TSRP-IUS-1-87, Trieste 1987.

\bibitem{wang} Y. Wang et al., ``Effect of surface roughness of optical elements on spatial
coherence of X-ray beams from third generation SR sources", Acta Optica Sinica 
20, 553-559 (2000)

\bibitem{snigirev:nima} A.~Snigirev, I.~Snigireva, V.~G.~Kohn, S.~M.~Kuznetsov,
``On the requirements to the instrumentation for the new generation of the synchrotron sources: berillium windows",
Nucl. Instrum.\& Meth. A, 370, pp 634-640 (1996)
On the requirements to the instrumentation for the new generation of the  radiation sources. Beryllium windows


\bibitem{snigirev:rsi} 	A. Snigirev, I. Snigereva, V. Kohn, S. Kuznetsov, 
I. Schelokov, ``On the possibilities of x-ray phase contrast microimaging by coherent high-energy synchrotron radiation,'' Rev. Sci. Instrum. 66, 5846-5492 (1995).


\bibitem{nugent:optexp}  K. A. Nugent, C. Q. Tran, and A. Roberts, ``Coherence transport through imperfect x-ray optical systems'', Optics Express Vol. 11, No. 19,  pp. 2323 - 2328 (2003).

\bibitem{vartanyants:optcom}	I.A. Vartanyants and I.K. Robinson, "Origins of decoherence in coherent X-ray diffraction experiments," Opt. Commun. 222, 29-50 (2003).

\bibitem{polak} F. Polack, D. Joyeux, J. Svatos, and D. Phalippou, ``Applications
of wavefront division interferometers in soft x rays,'' Rev. Sci.
Instrum. 66, 2180 (1995).


\bibitem{takayama} Takayama Y, Tai RZ, Hatano T, et al.
    Measurement of the coherence of synchrotron radiation
    J Synchrotron Radiat 5: 456-458 Part 3 MAY 1 1998 

\bibitem{xu} X.Xu et al.,
"Experimental investigation of spatial coherence for soft X-ray beam in
Hefei national SR facility", Acta Photonica Sinica 29, 29 (2000)

\bibitem{chang} C. Chang et al., "Spatial coherence characterization of undulator radiation",
Optics Comm. 182, 23-34 (2000)]

\bibitem{paterson}    Paterson D, Allman BE, McMahon PJ, et al.
    Spatial coherence measurement of X-ray undulator radiation
    Opt Commun 195 (1-4): 79-84 Aug 1 2001 

\bibitem{takayama1} Y. Takayama, T. Hatano,  T. Miyahara  and W. Okamoto
``Relationship Between Spatial Coherence of Synchrotron
Radiation and Emittance'', 
J. Synchrotron Rad. 5, 1187(1998).

\bibitem{marchesini:oe96}  S. Marchesini, R. Co\"\i sson, ``Two-dimensional coherence measurements with 
Fresnel mirrors", Opt. Eng. 35, 3597 (1996)

\bibitem{fezzaa} K. Fezzaa, F. Comin, S. Marchesini, R. Co\"\i sson and M. Belakhovsky, X-ray Interferometry using 2 coherent beams from Fresnel mirrors,  Journal of X-rays Science and Technology 7, 12-23, (1997)

\bibitem{marchesini} S. Marchesini, K. Fezzaa, M. Belakhovsky M, 
R. Co\"\i sson
X-ray interferometry of surfaces with Fresnel mirrors, Appl. Optics 39 (10)
 1633-1636, 2000.

\bibitem{leitenberger}    Leitenberger W, Kuznetsov SM, Snigirev A
    ``Interferometric measurements with hard X-rays using a double slit''
    Opt Commun 191 (1-2): 91-96 May 1 2001 

\bibitem{kohnprl}    Kohn V, Snigireva I, Snigirev A
    ``Direct measurement of transverse coherence length of hard x rays
from interference fringes'', Phys. Rev. Lett. 85 (13): 2745 (2000)

\bibitem{snigireva}    Snigireva I, Kohn V, Snigirev A
    Interferometric techniques for characterization of coherence of
high-energy synchrotron X-rays
    Nucl Instrum Meth A 467: 925-928 Part 2 Jul 21 2001 



\bibitem{cloetens}    Cloetens P, Guigay JP, DeMartino C, et al.
    Fractional Talbot imaging of phase gratings with hard x rays
    Opt Lett 22 (14): 1059-1061 JUL 15 1997 



\bibitem{lin:prl} J.~J.~A.~Lin, D. Paterson, A. G. Peele, P. J. McMahon, C. T. Chantler, and K. A. Nugent
``Measurement of the Spatial Coherence Function of Undulator Radiation using a Phase Mask'', Phys. Rev. Lett. 90, 074801 (2003) 

\bibitem{baron} A. Q. R. Baron ``Transverse coherence in nuclear resonant scattering of synchrotron radiation'' Hyperfine Interact 123 (1-8): 667-680, 1999 


\bibitem{yabashi:2001} M. Yabashi, K. Tamasaku, and T. Ishikawa, ``Characterization of the Transverse Coherence of Hard Synchrotron Radiation by Intensity Interferometry'', Phys. Rev. Lett. 87, 140801 (2001) 

\bibitem{yabashi:2002} M. Yabashi, K. Tamasaku, and T. Ishikawa, 
``Measurement of X-Ray Pulse Widths by Intensity Interferometry'', Phys. Rev. Lett. 88, 244801 (2002) 


\bibitem{abernathy}    Abernathy DL, Grubel G, Brauer S, et al.
    Small-angle X-ray scattering using coherent undulator radiation at
the ESRF    J Synchrotron Radiat 5: 37-47 Part 1 JAN 1 1998 


\bibitem{grubel} G. Gr\"ubel et al., ``Scattering with coherent X-rays", ESRF
Newsletter \underbar{20}, 14 (February 1994). 


\bibitem{schelekov} I.Schelokov, et al., ``X-ray interferometry
technique for mirror and multilayer characterisation", SPIE vol.2805, 282-292,
1996



\end{thebibliography}
\end{document}